\documentclass[aps,prc,showpacs,twocolumn,floatfix,superscriptaddress,nofootinbib]{revtex4}

\usepackage{amsmath,amssymb}

\usepackage{inputenc}
\usepackage{mathrsfs}
\usepackage{color}

\usepackage{epsfig,psfrag}
\usepackage{array}

\begin{document}

%\title{\boldmath {\slshape Ab initio} study of the $^{14}$F nucleus}
%\title{\boldmath Supercomputer %simulation 
%modeling  of {\slshape ab initio} nuclear
%structure: yet to be observed  $^{14}$F nucleus}

\title{\boldmath {\slshape Ab initio} nuclear
structure simulations:  the speculative $^{14}$F nucleus}

\author{P. Maris}
\affiliation{Department of Physics and Astronomy,
Iowa State University, Ames, IA 50011-3160, USA}
\author{A. M. Shirokov}
\affiliation{Skobeltsyn Institute of Nuclear Physics, Moscow State University,
Moscow, 119992, Russia}
\affiliation{Department of Physics and Astronomy,
Iowa State University, Ames, IA 50011-3160, USA}
\author{J. P. Vary}
\affiliation{Department of Physics and Astronomy,
Iowa State University, Ames, IA 50011-3160, USA}

\date{\today}

\begin{abstract}
We present results from {\em ab initio} No-Core Full Configuration
simulations of the exotic 
%neutron-deficient 
proton-rich
nucleus $^{14}$F whose
first experimental observation is expected soon.  The calculations
with JISP16 $NN$ interaction are performed up to the $N_{\max}=8$
basis space.  The binding energy is evaluated using an extrapolation
technique.  This technique is generalized to excitation energies, 
verified in calculations of $^6$Li and applied to $^{14}$B, the $^{14}$F mirror,
for which some data are available.
\end{abstract}

%\pacs{21.60.De, 21.60.Cs, 21.45.-v, 21.30.-x, 27.20.+n, 27.10.+h, 21.10.-k, 21.10.Dr}
\pacs{21.60.De, 27.20.+n, 21.10.Dr}
%, 21.45.-v} 
%
% 21.10.-k 	Properties of nuclei; nuclear energy levels
%  21.10.Dr 	Binding energies and masses 
% 21.30.-x 	Nuclear forces
% 21.45.-v 	Few-body systems
% 21.60.-n 	Nuclear structure models and methods
%  21.60.Cs 	Shell model 
%  21.60.De 	Ab initio methods
% 27.20.+h 	     A =< 5 
% 27.20.+n 	 6 =< A =< 19
% 27.20.+t 	20 =< A =< 38
% 27.20.+z 	39 =< A =< 58
%

\maketitle

%%%%%%%%%%%%%%%%%%%%%%%%%%%%%%%%%%%%%%%%%%%%%%%%%%%%%%%%%%%%%%%%%%%%%%%%%%%%%

Exotic nuclei at the nucleon drip lines and beyond constitute 
a forefront research area in nuclear physics.  The physics
drivers include: (1) to discover how shell structures evolve into 
extreme isospin regions; and (2) to extend our knowledge of 
the strong interactions, especially elusive three-nucleon forces 
(3NF), under these conditions.  In order to help pave a path
towards these goals, we present baseline {\it ab initio} results
for selected unstable A=13 and A=14 nuclei.

We focus especially on $^{14}$F, with
isospin T=2,  that is expected to lie beyond the proton
drip-line and therefore unstable.  This proton-rich nucleus will strain the
convergence properties of the {\em ab initio} methods we adopt here,
and also push the limits of state-of-the-art experimental facilities.
Indeed, the first experimental results regarding this four
proton excess isotope will be available soon from Cyclotron Institute
at Texas A\&M University \cite{Texas}.

We perform the first {\em ab initio} study of
$^{14}$F. We use the  No-Core Shell Model (NCSM) 
\cite{Vary,NCSM} which employs a many-body harmonic oscillator basis
which treats all nucleons as spectroscopically active.  The basis
space includes all many-body states with excitation quanta less then
or equal to $N_{\max}$ that makes it possible to completely remove
spurious center-of-mass excitations.  We used
the code MFDn \cite{Vary92_MFDn,ACM,SciDAC09} and the realistic $NN$
interaction JISP16\footnote{A Fortran code for the JISP16 interaction
  matrix elements  
is available at http://nuclear.physics.iastate.edu.} 
\cite{Shirokov:2005bk} which is known to
provide a good description of $p$ shell nuclei
\cite{Shirokov:2005bk,Maris:2008ax} without an additional 3NF.  
The largest calculations were
performed in the $N_{\max} = 8$ basis space, which for this nucleus
contains 1,990,061,078 basis states with total magnetic projection
$M=0$ and natural parity (negative).  The determination of the lowest
ten to fifteen eigenstates of the sparse Hamiltonian matrix,
for each oscillator parameter $\hbar\Omega$, requires 2 to 3 hours 
on 7,626 quad-core compute nodes at the Jaguar supercomputer at ORNL.

%%%%%%%%%%%%%%%%%%%%%%%%%%%%%%%%%%%%%%%%%%%%%%%%%%%%%%%%%%%%%%%%%%%%%%%%%%%%%
%\section{Ground state energy}

%
\begin{figure}[b]
  \epsfig{file=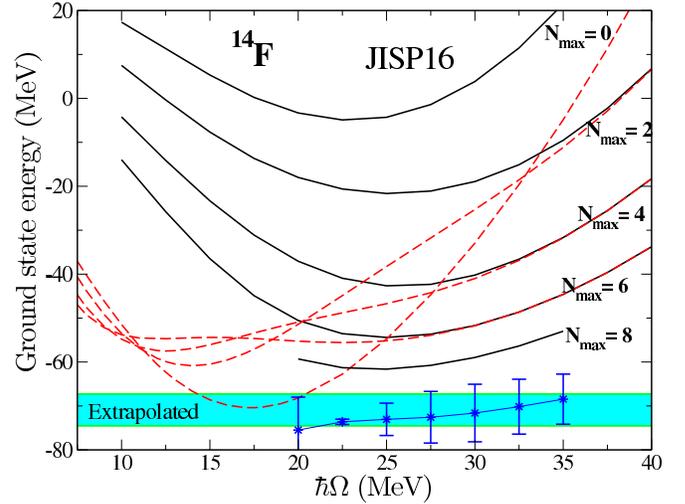,width=\columnwidth}
  \caption{\label{14Ff}
    (Color online) Results for the ground state energy of $^{14}$F
    with bare (solid) and LSO(2) renormalized (dashed) JISP16 as a
    function of the oscillator parameter $\hbar\Omega$.  The shaded
    area demonstrates the global extrapolation A for the binding
    energy and its uncertainty; the extrapolation B at fixed
    $\hbar\Omega$ is given by stars. The most reliable 
    $\hbar\Omega$ value for this extrapolation method 
    is at $\hbar\Omega=25$ MeV for $^{14}$F with 
    its uncertainty indicated by the error bar. }
\end{figure}
We show our complete results for the $^{14}$F ground state energy in
Fig.~\ref{14Ff}.   
The solid curves are our NCSM results with the
bare interaction JISP16.  These results are strict upper bounds for
the ground state energy, and converge monotonically with $N_{\max}$ to
the infinite basis space results.  The dashed curves in
Fig.~\ref{14Ff} are obtained in more conventional NCSM calculations
with effective $NN$ interactions derived from the initial bare
interaction JISP16 by the Lee--Suzuki--Okamoto (LSO) renormalization
procedure \cite{LSO}.  The renormalization procedure is truncated at
the two-body cluster level --- i.~e. induced three-body, four-body,
etc., contributions are neglected; hence we refer to these
calculations as LSO(2) renormalized.
Note that these results differ slightly from
preliminary approximate results presented at recent conferences
\cite{Vary:2008ay,Shirokov:2008ux}.

By comparing the bare and LSO(2) renormalized JISP16 results in
Fig.~\ref{14Ff}, we observe that the tendency of the LSO(2)
renormalized calculations is misleading.  For increasing basis spaces
from $N_{\max}=0$ to $6$, the minimum of the $\hbar\Omega$-dependent
curves increases, suggesting an approach from below to the infinite
basis space result.  At $N_{\max}=6$, LSO(2) renormalized JISP16
produces a nearly flat region at approximately the same energy as the
minimum obtained with the bare JISP16 interaction.  On the other hand,
the bare interaction provides a variational upper bound for the ground
state energy, which decreases with increasing $N_{\max}$.

Other light nuclei ($^6$He, $^6$Li, $^8$Be, $^{12}$C,
$^{16}$O, \ldots) show a qualitatively similar behavior: the LSO
renormalized interactions produce results which are neither an upper
bound nor a lower bound, and the approach to the infinite basis space
is non-monotonic.  Hence the convergence pattern of the LSO
renormalized results is difficult to assess.  Furthermore, with the
patterns displayed in Fig.~\ref{14Ff} for JISP16, the minima of the
$\hbar\Omega$-dependent ground state energy curves for both the bare
and the LSO(2) renormalized interaction may be expected to coincide
for $N_{\max} \ge 8$ as in some other nuclei.  For these reasons, we did
not perform expensive 
$N_{\max}=8$ LSO(2) renormalized JISP16 calculations for $^{14}$F.

Recently we introduced the {\em ab initio} No-Core Full Configuration
(NCFC) approach \cite{Maris:2008ax,Bogner:2007rx}, by extrapolating NCSM
results with the bare interaction in successive basis spaces to the
infinite basis space limit.  This makes it possible to obtain basis
space independent  results for binding energies and to evaluate
their numerical uncertainties.  We use two extrapolation methods: a
global extrapolation based on the  calculations in four successive
basis spaces and five $\hbar\Omega$ values in a 10 MeV interval
(extrapolation A); and extrapolation B based on the calculations
at various fixed $\hbar\Omega$ values in three successive basis spaces
and defining the most reliable $\hbar\Omega$ value for the
extrapolation.  These extrapolations provide consistent results
\cite{Maris:2008ax}.  Combining both extrapolation methods suggests a
binding energy of $72\pm 4$ MeV for $^{14}$F  which agrees
  well with AME03 nuclear binding energy extrapolations  \cite{AME03}, 
see Table~\ref{Egs}.
Ironically, out of all our NCSM calculations, both with the bare and
the LSO(2) renormalized interaction, the minimum of the LSO(2)
calculations at $N_{\max}=0$ appears to be closest to the infinite
basis space result.

\begin{table}[b]
  \caption{\label{Egs} 
    NCFC  results obtained with JISP16 for the ground state energies 
    (in MeV) of $^{13}$O, $^{14}$B and $^{14}$F.  Experimental 
    data are taken from Ref.~\cite{AjzenbergSelove:1991zz}.}
  \begin{ruledtabular}
    \begin{tabular}{l|ccc}
      Nucleus  &  Extrap. A   &  Extrap. B   &  Experiment \\ \hline
      $^{13}$O & $-75.7(2.2)$ & $-77.6(3.0)$ & $-75.56(0.01)$ \\
      $^{14}$B & $-84.4(3.2)$ & $-86.6(3.8)$ & $-85.42(0.02)$ \\
      $^{14}$F & $-70.9(3.6)$ & $-73.1(3.7)$ 
 &$-73.3(0.4)$\footnote{AME03 extrapolation \cite{AME03}.}
    \end{tabular}
  \end{ruledtabular}
\end{table}

To check the accuracy of our approach, we performed similar
calculations for the mirror nucleus $^{14}$B with a known binding
energy of 85.423 MeV \cite{AjzenbergSelove:1991zz}.  This value agrees
with our NCFC result of $86\pm 4$ MeV.  We also performed NCFC
calculations of the neighboring nucleus $^{13}$O using basis spaces up
to $N_{\max}=8$.  The calculated binding energy of $77\pm 3$ MeV also
agrees with the experimental value of 75.556 MeV
\cite{AjzenbergSelove:1991zz}.

We note that a good description of both $^{14}$F and $^{13}$O in the
same approach is important in order to have a consistent description
of the $^{13}{\rm O}+p$ reaction that produces $^{14}$F.  In this way,
any experimentally observed resonances can be directly compared with
the difference of our results for the $^{14}$F and $^{13}$O
energies.  In this respect it is interesting to note that although the
total energies of the extrapolations A and B differ by about 2 MeV,
the differences between the ground state energies of these three
nuclei are almost independent of the extrapolation method: for
$^{14}$F and $^{13}$O the predicted difference is 4.6 MeV, and for
$^{14}$F and $^{14}$B it is 13.5 MeV.  (The numerical uncertainty in
these differences is unclear, but expected to be significantly smaller
than the uncertainty in the total energies.)

%%%%%%%%%%%%%%%%%%%%%%%%%%%%%%%%%%%%%%%%%%%%%%%%%%%%%%%%%%%%%%%%%%%%%%%%%%%%%
%\section{Excitation spectrum}

We also calculated the $^{14}$F excitation spectrum in anticipation of
the experimental results.  It is unclear how to extrapolate excitation
energies obtained in finite basis spaces, but we can extrapolate the
total energies of excited states using the same methods as
discussed above for the ground state energy.  For the lowest state in
each $J^{\pi}$ channel the convergence pattern should be similar to
that of the ground state; for excited states with the same quantum
numbers we simply assume the same convergence pattern.  We perform
independent separate extrapolation fits for all states.  The
differences between the extrapolated total energies and the ground
state energy is our prediction for the excitation energies.

%%%%%%%%%%%%%%%%%%%%%%%%%%%%%%%%%%%%%%%%%%%%%%%%%%%%%%%%%%%%%%%%%%%%%%%%%%%%%
%\subsection{Spectrum of $^{6}$Li}

\begin{table*}[t]
  \caption{\label{t6Li}
 NCFC results for the $^{6}$Li ground state $E_{gs}$ and excitation
    energies $E_x$ (in MeV) obtained in different basis spaces with
    JISP16.  For extrapolations A and B we include in parentheses an
    estimate of the accuracy of the 
    %absolute binding 
    total
    energies; 
    for the LSO(2) renormalized interaction, we present the spread 
    in excitation energy for $\hbar\Omega$ variations 
    from 12.5 to 22.5 MeV.
    Experimental data are taken from Ref.~\cite{Tilley:2002vg}.}
  \begin{ruledtabular}    
    \begin{tabular}{lccc|ccc|cc}
      & Extrap. A  & Extrap. B  & LSO(2)
      & Extrap. A  & Extrap. B  & LSO(2)  &\multicolumn{2}{c}{Experiment}   \\ 
$E(J^\pi,T)$ 
      &  $N_{\max}=2{-}8$  &  $N_{\max}=4{-}8$  & $N_{\max}=6$ 
      & $N_{\max}=10{-}16$ & $N_{\max}=12{-}16$ & $N_{\max}=14$ 
      & Energy & Width  \\ \hline
$E_{gs}(1^+,0)_1$ 
      & $-30.9(0.6)$   & $-31.1(0.3)$   &       
      & $-31.47(0.09)$ & $-31.48(0.03)$ & 
      & $-31.994$ & Stable\\ \hline
$E_{x}(3^+,0)$ 
      & 2.6(0.5) & 2.5(1.2) & 2.2--2.7 %2.4(0.2) 
      & 2.56(0.04) & 2.55(0.07) & 2.53--2.55%2.55(0.07)
      & 2.186 & $24\cdot 10^{-3}$\\
$E_{x}(0^+,1)$ 
      & 3.6(0.6) & 3.5(1.2) & 3.3--3.7  %3.5(0.2)
      & 3.68(0.06) & 3.65(0.06) & %3.57--3.77 
                                  3.6--3.8 %3.7(0.1)
                                         & 3.563 &$8.2\cdot 10^{-6}$\\
$E_{x}(2^+,0)$ 
      & 5.3(0.9) & 5.5(1.8) & 4.8--5.8  %5.3(0.5)
      & 4.5(0.1) & 4.5(0.2) & 4.8-5.0 %4.8(0.2)
                                          & 4.312 & 1.30\\
$E_{x}(2^+,1)$ 
      & 6.3(0.7) & 6.1(1.6) & 6.2--6.5  %6.4(0.2)
      & 5.9(0.1) & 5.9(0.1) & 6.0--6.4 %6.2(0.2) 
                                          & 5.366 & 0.54\\
$E_{x}(1^+,0)_2$ 
      & 6.1(1.7) & 6.6(0.3) & 7.1--8.5  %7.8(0.7)
      & 5.4(0.3) & 5.4(0.2) & 6.1--6.6%6.3(0.5) 
                                          & 5.65 & 1.5
    \end{tabular}
  \end{ruledtabular}
\end{table*}

This approach to extrapolating the total eigenenergies is supported by
applying it to $^6$Li, see Table~\ref{t6Li}.  We have results for
$^6$Li in basis spaces up to $N_{\max} =16$ where a good convergence
is achieved and hence the extrapolation uncertainties are small.
These results are compared in Table~\ref{t6Li} with the extrapolations
based on calculations in basis spaces up to $N_{\max} = 8$, i.~e. in
the same basis spaces used for the $^{14}$F and $^{14}$B studies.

We see that the excitation energies based on $N_{\max} = 8$ and
smaller basis space results are consistent with the results obtained
in larger spaces.  The level ordering is the same and the difference
between the $N_{\max} = 8$ and $N_{\max} = 16$ results is generally
much smaller than the estimated uncertainties in the total energies
of the $N_{\max}= 8$ extrapolations.  This suggests that the numerical
uncertainty in the excitation energies is significantly smaller than
the uncertainty in the total energies: apparently, the calculated
total energies share a significant systematic uncertainty, an
overall binding  uncertainty, which
cancels when results are expressed as excitation energies.
Furthermore, we see that both extrapolation methods agree very well
with each other (within their error estimates), and that the error
estimates decrease as one increases the basis space.  

\begin{figure}
  \epsfig{file=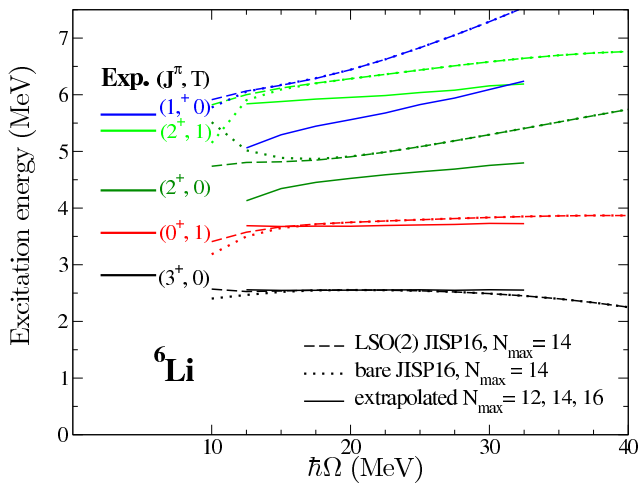,width=\columnwidth}
  \caption{\label{f6Li} 
    (Color online) NCSM results for the spectrum of $^{6}$Li with
    LSO(2) renormalized (dashed) and bare (dotted) JISP16 at 
    $N_{\max} = 14$, compared to NCFC extrapolations to infinite basis
    space (solid). 
% The most reliable $\hbar\Omega$ value for this
%  extrapolation method is at $\hbar\Omega=20$ MeV for $^{6}$Li.
    Experimental data are from Ref.~\cite{Tilley:2002vg}.}
\end{figure}

% new
The extrapolation B leads to results for the two lowest excited 
states that are practically independent of the oscillator parameter
$\hbar\Omega$, see Fig.~\ref{f6Li}.  Also the bare and LSO
renormalized NCSM results for these two states show very little
dependence on $\hbar\Omega$.  These states are narrow resonances, and
agree very well with experiment.
%

% new
On the other hand, the three higher excited states have a much larger
width, see Table~\ref{t6Li}.  Our calculations for these broad
resonances show a significant dependence on both $\hbar\Omega$ and
$N_{\max}$, in particular for the excited $(1^+,0)_2$ state which has
the largest width.  The extrapolation B to infinite model space
reduces but does not eliminate the $\hbar\Omega$ dependence.  We
further note that the $\hbar\Omega$-dependence of these excitation
energies is typical for wide resonances as observed in comparisons of
NCSM results with inverse scattering analysis of $\alpha$-nucleon
scattering states~\cite{Shirokov:2008jv}, and that the slope of the
$\hbar\Omega$ dependence increases with the width of the resonance.
This is consistent with the results presented in Fig.~\ref{f6Li}: the
width of the $(1^+,0)_2$ state is larger than the width of the
$(2^+,0)_2$ state; the latter is larger than the $(2^+,1)$ state
width.  Thus, there appears to be a significant correlation between
the resonance width and the $\hbar\Omega$ dependence.  The validity of
the extrapolation to infinite model space is not entirely clear for
these states.

We noted earlier that the LSO renormalized interaction does not
provide a monotonic approach to the infinite basis space for the
binding energies and this prevents simple extrapolation.  On the other
hand, the excitation energies with the LSO renormalized interaction
are often quite stable with $N_{\max}$.  However, it is important to
realize that this does not necessarily mean that these excitation
energies are numerically converged: they do depend on $\hbar\Omega$.
The dependence of the excitation energies on $\hbar\Omega$ decreases
slowly with increasing $N_{\max}$, as seen in
Table~\ref{t6Li}.  In fact, the excitation energies obtained with
LSO(2) renormalized JISP16 are nearly the same as those obtained with
the bare interaction, except at small values of $\hbar\Omega$, as
illustrated in Fig.~\ref{f6Li}.  For most states, the NCFC 
provides better results for the excitation
energies, with less basis space dependence than the LSO(2) NCSM
calculations in finite basis spaces.  Nevertheless, we can employ the
LSO procedure to obtain estimates of the binding and excitation
energies in small basis spaces where there are no NCFC results.

%%%%%%%%%%%%%%%%%%%%%%%%%%%%%%%%%%%%%%%%%%%%%%%%%%%%%%%%%%%%%%%%%%%%%%%%%%%%%
%\subsection{Spectra of $^{14}$F and $^{14}$B}

\begin{table*}
  \caption{\label{t14F-B}
 NCFC results for the $^{14}$F and $^{14}$B excitation energies $E_x$ (in MeV). 
    For extrapolations A and B we include in parentheses an
    estimate of the accuracy of the total  energies; 
    for the LSO(2) renormalized interaction, we present the spread 
    in excitation energy for $\hbar\Omega$ variations 
    from 12.5 to 22.5 MeV.  Experimental data are 
    taken from Ref.~\cite{Ball73}.
%\cite{AjzenbergSelove:1991zz}.
}
  \begin{ruledtabular}
    \begin{tabular}{lccc|ccc|cc}
      \multicolumn{7}{c|}{
	NCFC and NCSM {\it ab initio} calculations with JISP16}&\multicolumn{2}{c}{Experiment}\\
      \multicolumn{4}{c|}{$^{14}$F}&\multicolumn{3}{c|}{$^{14}$B} 
                                   & \multicolumn{2}{c}{$^{14}$B}\\
$E(J^\pi,T)$  & Extrap. A & Extrap. B & LSO(2)$,\:N_{\max}=6$
  &  Extrap. A & Extrap. B & LSO(2)$,\:N_{\max}=6$
            &$J^\pi$ &Energy  \\ \hline
$E_x(1^-,2)_1$  &0.9(3.9) & 1.3(2.5)  & 1.4--2.2 %1.8(0.4)
        & 1.1(3.5) & 1.4(2.8) &  1.4--2.3 %1.9(0.4)  
                                                    & $(1^-)$ 
%&0.74(4)  \\
&0.654(0.009)\footnote{Updated from Ref.~\cite{Kanungo05}.}\\
$E_x(3^-,2)_1$ & 1.9(3.3) & 1.5(4.6) & 1.0--1.8  %1.5(0.6) 
             &1.7(2.9) & 1.4(4.6) & 1.0--2.1 %1.5(0.6) 
                                                    & $(3^-)$ & 1.38(0.03)  \\
$E_x(2^-,2)_2$ & 3.2(3.5) & 3.3(3.5) &  3.3--3.7  %3.61(0.15)
    &3.3(3.1) & 3.3(3.8) & 3.5--3.8 %3.71(0.16)
                                                    & $2^-$ & 1.86(0.07) \\
$E_x(4^-,2)_1$ &3.2(3.2) & 2.8(4.8) & 2.0--2.6   %2.5(0.5)
      & 3.1(2.9) & 2.7(4.8) & 2.0--3.1 %2.5(0.5) 
                                                   & $(4^-)$ & 2.08(0.05)\\
& & & & & & & ? &[2.32(0.04)]\footnote{The
     existence of this state is uncertain (see Ref. \cite{Ball73}).}\\
& & & & & & & ? &2.97(0.04)\\
$E_x(1^-,2)_2$ & 5.9(3.5) & 5.4(4.6) & 5.8--6.4 %6.3(0.3) 
               & 5.9(3.1) & 5.5(4.8) & 5.7--6.4  %6.4(0.3)
                                                    \\
$E_x(0^-,2) $  & 5.1(5.4) & 5.8(1.0) & 5.8--10.5 %8.1(2.8)
               & 5.5(4.8) & 6.1(1.4) & 4.9--10.4 % 8.2(2.7) 
                                                   \\
$E_x(1^-,2)_3$ & 6.2(4.8) & 6.3(2.8) & 7.2--11.5 %9.1(2.7)
               & 6.4(4.3) & 6.4(3.1) & 6.1--11.3  %9.3(2.6)
                                                    \\
$E_x(2^-,2)_3$ & 6.4(4.6) & 6.3(3.4) & 7.3--10.9  %9.0(2.6)
               & 6.9(4.1) & 6.7(3.6) & 6.6--10.9  %9.2(2.2)
                                                    \\ 
$E_x(3^-,2)_2$ & 6.9(4.2) & 6.4(4.6) & 7.6--10.6 %8.9(2.1) 
               & 7.0(3.7) & 6.5(4.7) & 6.4--10.5 %9.0(2.0)
                                                    \\
$E_x(5^-,2)$ & 8.9(3.5) & 7.9(5.9) &  9.2--11.0 %10.0(1.2)
               & 8.8(3.1) & 7.8(5.9) & 8.5--10.8 %10.1(0.7)
%                                                    &\\
%\hline stop here? \\ \hline
%$E_x(2^-,2)_4$ %&5.8(6.8) 
%              %&7.2(3.7)
%              &7.7(4.6)&7.7(3.3) &10.0(4.7) &8.0(4.0) &7.9(3.6) & 10.2(4.7) &\\
%$E_x(3^-,2)_3$ &7.8(4.5) &7.6(3.5) &10.7(4.0) &8.3(4.0) &8.1(3.8) & 11.0(4.6)\\
%$E_x(4^-,2)_2$ &9.6(3.7) &12.3(5.4) &11.1(3.1) &9.5(3.4) & 9.0(4.7)& 11.1(2.9)
    \end{tabular}
  \end{ruledtabular}
\end{table*}

We summarize our results for the spectra of $^{14}$F and $^{14}$B in
Table~\ref{t14F-B}.  The excitation energies are obtained as a
difference between the extrapolated total energies of the excited
state and that of the ground state (see Table~\ref{Egs}).
The spectra are rather dense and the spacing between
energy levels is smaller than the quoted numerical uncertainty, which
is that of the extrapolated total energies of the excited states.
However, as discussed above, we expect that for narrow resonances the
actual numerical error in the excitation energy is (significantly)
smaller than the error in the total energy.
%, due to cancellations
%between errors in the total energies of the ground state and the
%excited states.

%
\begin{figure}
  \epsfig{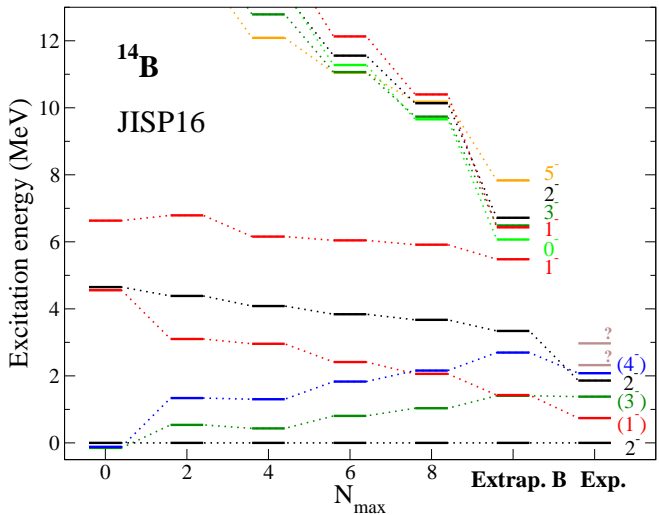}
  \caption{\label{14Bspectr}
    (Color online) Negative parity $^{14}$B spectrum obtained with
    JISP16 at fixed $\hbar\Omega=25$~MeV in successive basis spaces,
    and extrapolated to infinite basis space using Extrapolation B.
    Experimental data are taken from Ref.~\cite{AjzenbergSelove:1991zz}.}
\end{figure}
Figure~\ref{14Bspectr} shows that different excited states can have
very different convergence behavior.  (Although we presented in
Fig.~\ref{14Bspectr} the $^{14}$B results, the behavior of the
$^{14}$F states is similar.)  At $N_{\max}=8$, there are five
low-lying excited states; the excitation energy of these states
depends only weakly on the basis space as $N_{\max}$ increases from
$2$ to $8$.  Then there are numerous higher excited states which
depend strongly on the basis space: their excitation energies decrease
rapidly with increasing $N_{\max}$.  Only after extrapolation to the infinite 
basis space do they appear at excitation energies comparable
to the other low-lying excited states.  We see a similar phenomenon in
NCFC calculations of other nuclei.

\begin{figure}
  \epsfig{file=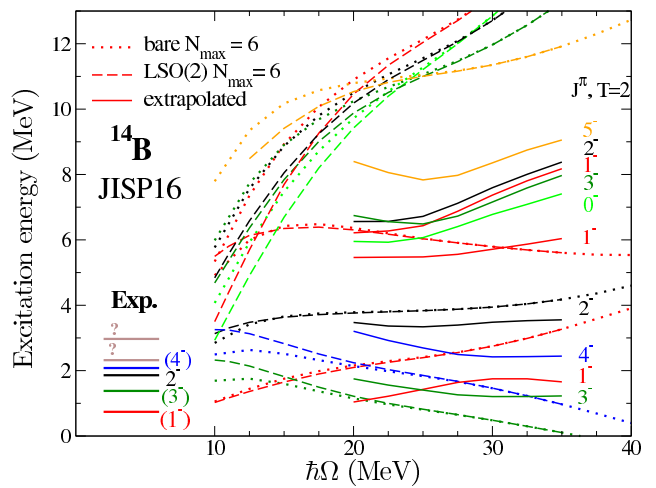,width=\columnwidth}
  \caption{\label{14BhW} 
    (Color online) NCSM results for the negative parity spectrum of
    $^{14}$B with LSO(2) renormalized (dashed) and bare (dotted)
    JISP16 at $N_{\max} = 6$, compared to NCFC extrapolations to
    infinite basis space (solid) from $N_{\max}=4{-}8$.  
    The most reliable $\hbar\Omega$ value for this extrapolation 
    method is at $\hbar\Omega=25$ MeV for all states depicted.  
    Experimental data are taken from
    Ref.~\cite{AjzenbergSelove:1991zz}.}
\end{figure}
The dependence on $\hbar\Omega$ varies considerably over
the excited states as seen in  Fig.~\ref{14BhW}.
The lowest five excited states have a weak dependence on $\hbar\Omega$, 
whereas the higher excited states depend
strongly on it.  We expect our results for these higher
excited states to have a larger numerical error than our results for
the lower excited states with the weaker $\hbar\Omega$ dependence.
Furthermore, in analogy to the excited states in $^{6}$Li
discussed above, we expect these higher states to be broad resonances.
Interestingly, the high-lying $J^{\pi}=5^-$ state has a relatively
weak $\hbar\Omega$ dependence (compared to states with similar
excitation energy); it  is also less dependent on $N_{\max}$, and may
correspond to a narrower resonance.

{ Note, the conventional wisdom suggests leading
configurations for the
ground and 5 lowest-lying levels of $^{14}$F ($^{14}$B) to be formed by
the $p_{3/2}$ neutron (proton) and $s_{1/2}$ or $d_{5/2}$ proton
(neutron). Other low-lying states, with the exception of our low-lying
$5^-$ state, involve $p_{1/2}$ and/or $d_{3/2}$ single-particle states.
}

In Fig.~\ref{14BhW} we can also see that the excitation energies obtained
with LSO(2) renormalized JISP16 are nearly the same as those obtained
with the bare interaction, at least at $N_{\max}=6$.  Note that the
NCFC results differ significantly from the bare and LSO(2) results, in
particular for the higher excited states with a strong $N_{\max}$
dependence; these extrapolated results also tend to have a somewhat
weaker dependence on $\hbar\Omega$ than the results in finite basis
spaces, and are  expected to be more accurate.

Some of the excited states in $^{14}$B were observed experimentally.
Unfortunately, the spin of most of these states is doubtful or unknown.  Overall,
the calculated excitation energies appear to be too large when compared
with the experimental data; in particular our result for the
excited $2^-$ state, the only excited state with a firm spin
assignment, is about 1.5 MeV above the experimental value.  However,
the spin of the lowest five states agrees with experiment, except for
the $2^-$ and $4^-$ being interchanged, assuming that the tentative
experimental spin assignments are correct.  We do not see additional
states between 2 and 3 MeV, but this could be related to the fact that
all our excitation energies appear to be too large.  Also, given the
strong dependence on $N_{\max}$ of the higher excited states, it is
not unlikely that these states will come down further with increasing
basis space.

%%%%%%%%%%%%%%%%%%%%%%%%%%%%%%%%%%%%%%%%%%%%%%%%%%%%%%%%%%%%%%%%%%%%%%%%%%%%%
%\section{Summary}

We performed the first theoretical {\em ab initio} study of the exotic 
proton-rich nucleus $^{14}$F which has yet to be observed
experimentally.  Using the $NN$ interaction JISP16, we presented a
prediction for the $^{14}$F binding energy that is supported by
comparing our NCFC results with experimental data for the binding
energies of the mirror nucleus $^{14}$B and the neighboring nucleus
$^{13}$O obtained within the same approach in the same basis spaces.

We extended our NCFC extrapolation techniques to evaluate excited
states, and validated this method by applying it to excited states in
$^6$Li.  The obtained spectrum for $^{14}$B agrees qualitatively
with the limited data, and we made predictions for the spectrum of
$^{14}$F.  More definite information about the excited states in
$^{14}$B would be helpful.  It would also be very interesting to
compare our predictions for the $^{14}$F binding energy and spectrum
with the experimental data that are anticipated soon.  Significant
differences between our predictions and the experimental results would
indicate deficiencies in the $NN$ interaction, JISP16, and/or the role
of neglected 3NF's.  This would inform future research efforts and,
with the inclusion of additional unstable nuclei in the analysis, aid
in the eventual determination of the underlying shell structure
evolution.

% modified
Although NCSM calculations with LSO(2) renormalized interactions
generally give reasonable results for the binding energies and spectra
in small basis spaces, they do not improve systematically with
increasing basis space.  In particular for JISP16 we find that the
results for the bare and the LSO(2) renormalized interaction basically
coincide for $N_{\max} \ge 8$, both for total energies and for
excitation energies.  It would be worthwhile, although it is a major
undertaking, to evaluate the effects of the induced three- and
four-body interactions, which should improve the accuracy of the LSO
renormalized calculations.  Without a thorough study of those effects
however, we prefer the NCFC approach, based on extrapolations of NCSM
results with the bare interaction, at least for JISP16.

%%%%%%%%%%%%%%%%%%%%%%%%%%%%%%%%%%%%%%%%%%%%%%%%%%%%%%%%%%%%%%%%%%%%%%%%%%%%%
%\section{Acknowledgements}

We thank V.~Z.~Goldberg (Texas A\&M University) for very valuable
discussions.  We also thank Esmond Ng, Chao Yang and Philip Sternberg
of LBNL and Masha Sosonkina of Ames Laboratory for fruitful
discussions on computational science and applied mathematics issues
underlying code developments.  This work was supported by the US DOE
Grants DE-FC02-09ER41582 and DE-FG02-87ER40371 and the 
%FAO 
Russian Federal Agency of Education Contract
P521.  Computational resources were provided by DOE through the
National Energy Research Supercomputer Center (NERSC) and through an
INCITE award (David Dean, PI); the $^6$Li runs where performed on the
Franklin supercomputer at NERSC and the $^{14}$F and $^{14}$B runs on
Jaguar at ORNL.

%%%%%%%%%%%%%%%%%%%%%%%%%%%%%%%%%%%%%%%%%%%%%%%%%%%%%%%%%%%%%%%%%%%%%%%%%%%%%

\end{document}